\documentclass[a4paper,10pt]{article}

\usepackage{contribution}
\newcommand{\weblink}[2][]{%
    \ifthenelse{\equal{#1}{}}%
    {\textnormal{\url{#2}}}%
    {\textnormal{\href{#2}{#1}}}%
}

\def\beq{\begin{equation}}
\def\eeq#1{\label{#1}\end{equation}}
\def\eeqn{\end{equation}}

\def\beqa{\begin{eqnarray}}
\def\eeqa#1{\label{#1}\end{eqnarray}}
\def\eeqan{\end{eqnarray}}

\let\bar=\overbar

\def\Dslash{\not{\hbox{\kern-4pt $D$}}}
\def\dslash{\not{\hbox{\kern-2pt $\del$}}}

\def\msb{{\bar{\ssstyle M \kern -1pt S}}}

\newcommand{\contribution}[7][]{%
  \clearpage
  \thispagestyle{plain}

  \ifthenelse{\equal{#1}{}}
  {\hypersetup{pdftitle={#2}}}
  {\hypersetup{pdftitle={#1}}}
  \hypersetup{pdfauthor={{#3} {#4}}}
  {\centering\normalfont\LARGE\bfseries\sffamily #2 \par\nobreak}
  \lhead{}
  \chead{%
    \textit{\footnotesize XXIInd International Workshop ``High-Energy Physics and Quantum Field Theory'', 
June 24 -- July 1, 2015, Samara, Russia}%
  }
  \rhead{}
  \bigskip
  \begin{center}
    {#3} {#4}\ifthenelse{\equal{#6}{}}{}{\footnote{\weblink[#6]{mailto:#6}}}
    \ifthenelse{\equal{#7}{}}{}{#7} \\
    \textit{#5}
  \end{center}
  \bigskip
}

\renewcommand{\abstract}[1]{%
  \begin{center}
    \begin{minipage}{0.85\textwidth}
      \begin{footnotesize}
        #1
      \end{footnotesize}
    \end{minipage}
  \end{center}
  \bigskip
}

\begin{document} 

% % % % % % % % % % % % % % % % % % % % % % % % % % % % % % % % % % % % % % % % %
% your proceedings
% template for qfthep2015 contribution
%
% please do not rename this file
%
% to create document run
%
%     pdflatex qfthep (if you have pdf figures)
%
%     or 
%     latex qfthep
%     dvips qfthep
%     ps2pdf qfthep.ps
%
%     Finally, please rename your contribution qfthep.pdf -> <your name>.pdf 
%     and send us <your name>.pdf file
%
%%%%%%%%%%%%%%%%%%%%%%%%%%%%%%%%%%%%%%%%%%%%%%%%%%%%%%%%%%%%%%%%%%%%%%%%%%%%%%%%%
{  % do not remove
%%%%%%%%%%%%%%%%%%%%%%%%%%%%%%%%%%%%%%%%%%%%%%%%%%%%%%%%%%%%%%%%%%%%%%%%%%%%%%%%%
% template for articles submitted to the full-tex econf proceedings

%%%%%%%%%%%%%%%%%%%%%%%%%%%%%%%%%%%%%%%%%%%%%%%%%%%%%%%%%%%%%%%%%%%%%%%%%%%%%%%%%
% please define your macros here

%
%%%%%%%%%%%%%%%%%%%%%%%%%%%%%%%%%%%%%%%%%%%%%%%%%%%%%%%%%%%%%%%%%%%%%%%%%%%%%%%%%

%%%%%%%%%%%%%%%%%%%%%%%%%%%%%%%%%%%%%%%%%%%%%%%%%%%%%%%%%%%%%%%%%%%%%%%%%%%%%%%%%
% define title, author, and address
% contribution[short title]{title}{author first name}{author last name}{author address}{author email}{collaboration}
% the short title will appear in the page headers and the TOC of the book of proceedings
% the last two arguments may be left empty
\contribution[Weak-annihilation rare radiative decays of B-mesons]  % short title (optional)
{Weak-annihilation rare radiative decays of B-mesons}  % title
{Anastasiia}{Kozachuk}  % first and last name of author
{D.~V.~Skobeltsyn Institute of Nuclear Physics, \\
M.~V.~Lomonosov Moscow State University\\
Leninskie Gory 1, 119991 Moscow, Russia}  % author address
{anastasiia.kozachuk@cern.ch}  % author email optional anastasiia.kozachuk@cern.ch
{}  % collaboration (optional, co-authors appear here )
%
%%%%%%%%%%%%%%%%%%%%%%%%%%%%%%%%%%%%%%%%%%%%%%%%%%%%%%%%%%%%%%%%%%%%%%%%%%%%%%%%%

%%%%%%%%%%%%%%%%%%%%%%%%%%%%%%%%%%%%%%%%%%%%%%%%%%%%%%%%%%%%%%%%%%%%%%%%%%%%%%%%%
% abstract
\abstract{%
 In this talk, we present the predictions \cite{Kozachuk:2015kos} for a number of radiative decays $B_{(s)}\to V\gamma$, with the vector meson in the final state, which proceed through the weak-annihilation mechanism. Within the factorization approximation, we take into account the photon 
emission from the $B$-meson loop and from the vector-meson loop. The highest branching ratios for the weak-annihilation reactions reported here are 
${\cal B}(\bar B^0_s\to J/\psi\gamma)=1.5\cdot 10^{-7}$ and  
%${\cal B}(\bar B^0_d\to J/\psi\gamma)=8\cdot 10^{-9}$, 
%${\cal B}(\bar B^0_d\to D^{0*}\gamma)=4.5\cdot 10^{-8}$, 
${\cal B}(B^-\to \bar D_s^{*-}\gamma)=1.7\cdot 10^{-7}$, 
the estimated accuracy of these predictions being at the level of 20\%. 
}
%
%%%%%%%%%%%%%%%%%%%%%%%%%%%%%%%%%%%%%%%%%%%%%%%%%%%%%%%%%%%%%%%%%%%%%%%%%%%%%%%%%

%%%%%%%%%%%%%%%%%%%%%%%%%%%%%%%%%%%%%%%%%%%%%%%%%%%%%%%%%%%%%%%%%%%%%%%%%%%%%%%%%
% main text
% for short contributions sections are optional
\section{Introduction}
The investigation of rare $B$ decays forbidden at the tree level in the Standard Model provides the 
possibility to probe the electroweak sector at large mass scales. Interesting information 
about the structure of the theory is contained in the Wilson coefficents entering the 
effective Hamiltonian which take different values in different theories with testable 
consequences in rare $B$ decays. 

There is an interesting class of rare radiative $B$-decays which proceed merely through the weak-annihilation mechanism. 
These processes have very small probabilities and have not been observed. 
So far, only upper limits on the branching ratios of these decays have been obtained: In 2004, the BaBar Collaboration 
provided the upper limit ${\cal B} (B^0\to J/\psi\gamma)<1.6\cdot 10^{-6}$ \cite{BaBar2004}. 
Very recently, the LHCb Collaboration reached the same sensitivity to the $B^0$-decay and set the limit on the $B_s^0$ decay: 
${\cal B} (B^0\to J/\psi\gamma)<1.7\cdot 10^{-6}$ and ${\cal B} (B_s^0\to J/\psi\gamma)<7.4\cdot 10^{-6}$ at 90\% CL 
\cite{LHCb2015}. 
Obviously, with the increasing statistics, the prospects to improve the limits on the branching ratios by one order of magnitude 
or eventually to observe these decays in the near future seem very favourable. 

The annihilation-type $B$-decays are promising from the perspective of obtaining theoretical predictions since the QCD dynamics 
of these decays is relatively simple \cite{gp,grinstein}. These decays have been addressed in the literature but --- in spite 
of their relative 
simplicity --- the available theoretical predictions turned out to be rather uncertain; for instance, the predictions for 
${\cal B} (B_s^0\to J/\psi\gamma)$ decay vary from $5.7\cdot 10^{-8}$ \cite{ch2004} to $5\cdot 10^{-6}$ \cite{ch2006}. 
The situation is clearly unsatisfactory and requires clarification. We did not find any of these results convincing and 
present in this paper a more detailed analysis of the $B\to V\gamma$ decays. 

The annihilation type $B\to V\gamma$ decays proceed through the four-quark operators of the effective weak Hamiltonian. 
In the factorization approximation, the amplitude can be represented as the product of meson leptonic decay constants 
and matrix elements of the weak current between meson and photon; the latter contain the meson-photon transition form factors. 
The photon can be emitted from the loop containing the $B$-meson (Fig.~\ref{fig:diag}a), this contribution is described by 
the $B\gamma$ transition form factors. The photon can be also emitted from the vector-meson $V$-loop (Fig \ref{fig:diag}b); 
this contribution is described by the $V\gamma$ transition form factors. The latter were erroneuosly believed to give 
small contribution to the amplitude and have not been considered in the previous analyses.  
\begin{figure}[h!]
\centering
\begin{tabular}{c}
\includegraphics[width=10cm]{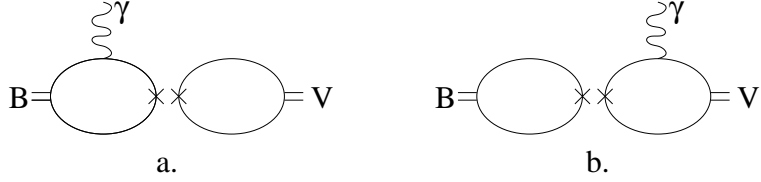}  
\end{tabular}
\caption{\label{fig:diag}
Diagrams describing the weak annihilation process for $B\to V\gamma$ in the factorization approximation: 
(a) The photon is emitted from the $B$-loop, 
(b) The photon is emitted from the vector-meson $V$-loop.}
\end{figure}

We calculate the $B\gamma$ and $V\gamma$ form factors within the relativistic dispersion approach based on 
the constituent quark picture 
\cite{m}. As shown in  \cite{melikhov}, the form factors from this approach satisfy all rigorous constrains which 
emerge in QCD in the 
limit of heavy-to-heavy and heavy-to-light transitions; as demonstrated in \cite{ms,bmns,mn2004}, the numerical results for the 
weak transition form factors from this approach exhibit an excellent agreement with the results from lattice QCD and 
QCD sum rules. 

%===========================================================================================================
\section{\label{Sect2}The effective Hamiltonian, the amplitude, and the decay rate}
We consider the weak-annihilation radiative $B\to V\gamma$ transition, where 
$V$ is the vector meson containing at least one charm quark, i.e. having the quark content $\bar q c$ ($q=u,d,s,c$). 
The corresponding amplitude is given by the matrix element of the effective Hamiltonian \cite{heff}
\begin{eqnarray}
A(B\to V\gamma)=\langle \gamma(q_1)V(q_2)|H_{\rm eff}|B(p) \rangle,  
\end{eqnarray}
where $p$ is the $B$ momentum, $q_2$ is the vector-meson momentum, and $q_1$ is the photon
momentum, $p=q_1+q_2$, $q_1^2=0$, $q_2^2=M_V^2$, $p^2=M_B^2$. 
The effective weak Hamiltonian relevant for the transition of interest has the form:
\begin{eqnarray}
\label{Heff}
H_{\rm eff} &=& 
-\frac{G_F}{\sqrt{2}}{\xi_{\rm CKM}}
\left (C_1(\mu){\cal O}_1+C_2(\mu){\cal O}_2\right), 
\end{eqnarray}
$G_F$ is the Fermi constant, $\xi_{\rm CKM}=V^*_{cd}V_{ub}$, $C_{1,2}(\mu)$ are the scale-dependent Wilson coefficients \cite{heff}, 
and we only show the relevant four-quark operators 
\begin{eqnarray}
{\cal O}_1 &=& \bar d_{\alpha}\gamma_{\nu}(1-\gamma_5)c_{\alpha}\;
\bar u_{\beta}\gamma_{\nu}(1-\gamma_5) b_{\beta},\nonumber
\\
{\cal O}_2 &=& \bar d_{\alpha}\gamma_{\nu}(1-\gamma_5) c_{\beta}\; 
\bar u_{\beta}\gamma_{\nu}(1-\gamma_5) b_{\alpha}.
%\nonumber\\
\end{eqnarray}
We use notations $e=\sqrt{4\pi\alpha_{\rm em}}$, 
$\gamma^5=i\gamma^0\gamma^1\gamma^2\gamma^3$,  
$\sigma_{\mu\nu}=i\left [\gamma_{\mu},\gamma_{\nu}\right ]/2$,
$\epsilon^{0123}=-1$ and
${\rm Sp}\left (\gamma^5\gamma^{\mu}\gamma^{\nu}\gamma^{\alpha}\gamma^{\beta}\right )
=4i\epsilon^{\mu\nu\alpha\beta}$.

In the amplitude it is convenient to isolate the parity-conserving contribution which emerges from the
product of the two equal-parity currents, and the parity-violating contribution 
which emerges from the product of the two opposite-parity currents. The amplitude may then be parametrized as follows  
\begin{eqnarray}
\label{F_PC}
A(B\to V\gamma)=\frac{eG_F}{\sqrt{2}}
\left[
\epsilon_{q_1\epsilon^\ast_1 q_2 \epsilon_2^\ast}F_{\rm PC}
+i \epsilon_2^{\ast\nu}\epsilon_1^{\ast\mu} \left(g_{\nu\mu}\,pq_1-p_\mu q_{1\nu}\right)F_{\rm PV}
\right], 
\end{eqnarray}
where $F_{\rm PC}$ and $F_{\rm PV}$ are the parity-conserving and 
parity-violating invariant amplitudes, respectively. Hereafter $\epsilon_2$($\epsilon_1$) is the 
vector-meson (photon) polarization vector. We use the short-hand notation 
$\epsilon_{abcd}=\epsilon_{\alpha\beta\mu\nu}a^{\alpha}b^{\beta}c^{\mu}d^{\nu}$ 
for any 4-vectors $a,b,c,d$. 

For the decay rate one finds 
\begin{eqnarray}
\label{rate}
\Gamma(B\to V\gamma)=\frac{G^2_F\,\alpha_{em}}{16}M_B^3
\left(1-{M^2_V}/{M_B^2}\right)^3
     \left( |F_{\rm PC}|^2+|F_{\rm PV}|^2 \right). 
\end{eqnarray}

After neglecting the nonfactorizable soft-gluon exchanges, i.e. assuming vacuum saturation, it is convinient to parametrize the parity-violating and the parity-conserving amplitudes of (\ref{F_PC}) in the following way
\begin{eqnarray}
F_{\rm PV}&=&\xi_{\rm CKM}a_{\rm eff}(\mu)\left[\frac{F_{A}}{M_B}f_V M_V+f_B H_S
-\frac{2Q_Bf_Bf_VM_V}{M_B^2-M_V^2}\right], \\
\label{fpc}
F_{\rm PC}&=&\xi_{\rm CKM} a_{\rm eff}(\mu)\left[\frac{F_V}{M_B}f_VM_V +f_B H_P\right]. 
\end{eqnarray}

Summing up this Section, within the factorization approximation the weak annihilation amplitude 
can be expressed in terms of four form factors: $F_A$, $F_V$, $H_P$ and $H_S$. 
It should be emphasized that each of the form factors $F_A$, $F_V$, $H_P$ and $H_S$ actually depends on two variables: 
The $B$-meson transition form factors $F_A$, $F_V$ depend on $q_1^2$ and $q_2^2$, and  
$F_{A,V}(q_1^2,q_2^2)$ should be evaluated at $q_1^2=0$ and $q_2^2=M_V^2$. 
The vector-meson transition form factors $H_P$ and $H_S$ 
depend on $q_1^2$ and $p^2$, and $H_{S,P}(q_1^2,p^2)$ should be evaluated at $q_1^2=0$ and $p^2=M_B^2$.

%=========================================================================================================
\section{\label{Sect3}Photon emission from the $B$-meson loop and the form factors $F_A$ and $F_V$.}
In this section we calculate the form factors $F_{A,V}$ within the relativistic dispersion approach 
to the transition form factors based on constituent quark picture. This approach has been formulated in detail in 
\cite{melikhov} and applied to the weak decays of heavy mesons in \cite{ms}. 

The pseudoscalar meson in the initial state is described in the dispersion approach by the following vertex \cite{m}: 
$\bar q_1(k_1)\; i\gamma_5 q(-k_2)\;G(s)/{\sqrt{N_c}}$, 
with $G(s)=\phi_P(s)(s-M_P^2)$, $s=(k_1+k_2)^2$, $k_1^2=m_1^2$ and $k_2^2=m_2^2$. 
The pseudoscalar-meson wave function $\phi_P$ is normalized according to the relation \cite{m}
\begin{eqnarray}
\label{norma}
\frac{1}{8\pi^2}\int\limits_{(m_1+m_2)^2}^\infty ds \phi_P^2(s)
\left({s-(m_1-m_2)^2}\right)\frac{\lambda^{1/2}(s,m_1^2,m_2^2)}{s}=1.  
\end{eqnarray}
The decay constant is represented through $\phi_P(s)$ by the spectral integral
\begin{eqnarray}
\label{fP}
f_P=\sqrt{N_c}\int\limits_{(m_1+m_2)^2}^\infty ds \phi_P(s)
(m_1+m_2)\frac{\lambda^{1/2}(s,m_1^2,m_2^2)}{8\pi^2s}\frac{s-(m_1-m_2)^2}{s}. 
\end{eqnarray}
Here $\lambda(a,b,c)=(a-b-c)^2-4bc$ is the triangle function. 

The form factor $F_A$ is given by the diagrams of Fig \ref{fig:Fa}. 
Fig \ref{fig:Fa}a shows $F_A^{(b)}$, the contribution to the form factor of the process when the $b$ quark interacts with the 
photon; Fig \ref{fig:Fa}b describes the contribution of the process when the quark $u$ interacts while 
$b$ remains a spectator. 

It is convenient to change the direction of the quark line in the loop diagram of 
Fig \ref{fig:Fa}b. This is done by performing the charge conjugation of the matrix element 
and leads to a sign change for the $\gamma_\nu\gamma_5$ vertex, so that both diagrams of Fig \ref{fig:Fa} 
can reduced to one diagram. The same is done for the $F_V$ form factor. The resulting diagrams are represented on Fig \ref{fig:Fat}:

\begin{figure}[h!]
\centering
\begin{tabular}{c}
\includegraphics[width=9cm]{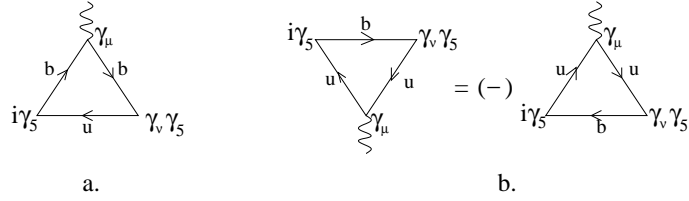} 
\end{tabular}
\caption{\label{fig:Fa}Diagrams for the form factor $F_A$: a) $F_A^{(b)}$, b) $F_A^{(u)}$.} 
\end{figure}

\begin{figure}[h!]
\centering
\begin{tabular}{cccc}
\includegraphics[width=3cm]{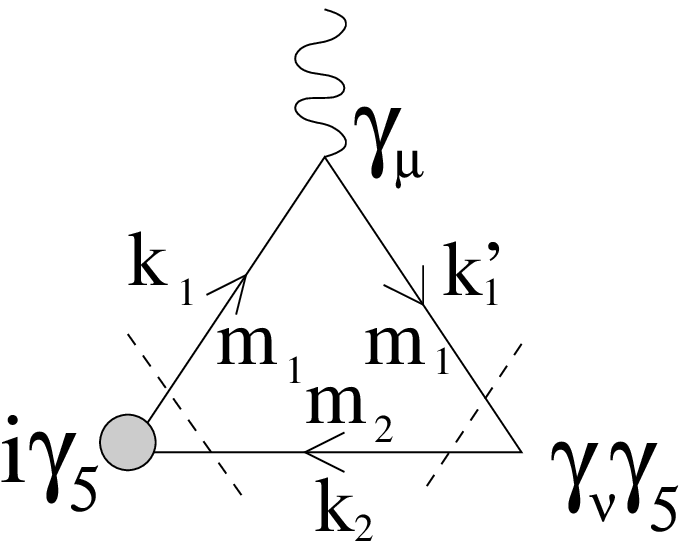} &&&
\includegraphics[width=3cm]{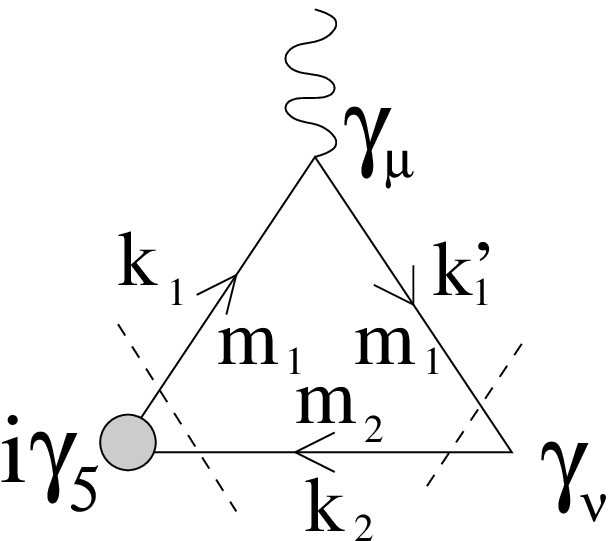} 
\end{tabular}
\caption{\label{fig:Fat} 
The triangle diagram for $F_A^{(1)}(m_1,m_2)$ and $F_V^{(1)}(m_1,m_2)$ form. The cuts correspond to calculating the double spectral 
density in $p^2$ and $q_2^2$.}
\end{figure} 

Setting $m_1=m_b$, $m_2=m_u$ gives $F_A^{(b)}$ and $F_V^{(b)}$, while 
setting $m_1=m_u$, $m_2=m_b$ gives $F_A^{(u)}$ and $F_V^{(u)}$ such that      
\begin{eqnarray}
\begin{array} {lll}
F_A=Q_b F_A^{(b)}-Q_u F_A^{(u)}, && F_V=Q_bF_V^{(b)}+Q_uF_V^{(u)}. 
\end{array}
\end{eqnarray} 

This expression may be cast in the form of a single dispersion integral 
\begin{eqnarray}
\label{fadisp}
\frac{1}{M_B}F_{A}^{(1)}(m_1,m_2)&=&\frac{\sqrt{N_c}}{4\pi^2}\int\limits_{(m_1+m_2)^2}^\infty
\frac{ds\;\phi_B(s)}{(s-M_V^2)}
\left(\rho_+(s,m_1,m_2)+2\frac{m_1-m_2}{M_B^2-M_V^2}\rho_{k_\perp^2}(s,m_1,m_2)\right), \\ 
\frac{1}{M_B}F_V^{(1)}(m_1,m_2)&=&-\frac{\sqrt{N_c}}{4\pi^2}\int\limits_{(m_1+m_2)^2}^\infty
\frac{ds\phi_B(s)}{(s-M_V^2)}\rho_+(s,m_1,m_2). 
\end{eqnarray}

where 
\begin{eqnarray}
\label{rhoplus}
\rho_+(s,m_1,m_2)&=&(m_2-m_1)\frac{\lambda^{1/2}(s,m_1^2,m_2^2)}{s}+m_1\log\left(\frac{s+m_1^2-m_2^2+\lambda^{1/2}(s,m_1^2,m_2^2)}
{s+m_1^2-m_2^2-\lambda^{1/2}(s,m_1^2,m_2^2)}\right),
\\
\label{rhokperp2}
\rho_{k_\perp^2}(s,m_1,m_2)&=&\frac{s+m_1^2-m_2^2}{2s}\lambda^{1/2}(s,m_1^2,m_2^2)-
m_1^2\log\left(\frac{s+m_1^2-m_2^2+\lambda^{1/2}(s,m_1^2,m_2^2)}
{s+m_1^2-m_2^2-\lambda^{1/2}(s,m_1^2,m_2^2)}\right).
\end{eqnarray}

%=========================================================================================================
\section{\label{Sect4}Photon emission from the vector meson loop. The form factors $H_S$ and $H_P$.}
We now calculate the form factors $H_{P,S}$ using the relativistic dispersion approach. 
The vector meson in the final state is described in this approach by the vertex $\bar q_2(-k_2)\Gamma_\beta q_1(k_1')$, 
$\Gamma_\beta=\left(-\gamma_\beta+\frac{(k_1'-k_2)_\beta}{\sqrt{s}+m_1+m_2}\right)\;G(s)/{\sqrt{N_c}}$, 
with $G(s)=\phi_V(s)(s-M_V^2)$, $s=(k'_1+k_2)^2$, ${k'}_1^2=m_1^2$ and $k_2^2=m_2^2$. 
The vector-meson wave function $\phi_V$ is normalized according to \cite{melikhov}
\begin{eqnarray}
\label{normaV}
\frac{1}{8\pi^2}\int\limits_{(m_1+m_2)^2}^\infty ds \phi_V^2(s)
\left({s-(m_1-m_2)^2}\right)\frac{\lambda^{1/2}(s,m_1^2,m_2^2)}{s}=1.  
\end{eqnarray}
Its decay constant is represented through $\phi_V(s)$ by the spectral integral
\begin{eqnarray}
\label{fV}
f_V=\sqrt{N_c}\int\limits_{(m_1+m_2)^2}^\infty ds \phi_V(s)
\frac{2\sqrt{s}+m_1+m_2}{3}\frac{\lambda^{1/2}(s,m_1^2,m_2^2)}{8\pi^2s}\frac{s-(m_1-m_2)^2}{s}. 
\end{eqnarray}
Now the form factors $H_{S,P}$ describe the transition of the current with momentum $p$, $p^2=M_B^2$, 
to the photon with momentum 
$q_1$, $q_1^2=0$, and the vector meson with the momentum $q_2$, $q_2^2=M_V^2$. 
Similar to the previous section, we derive the double spectral representations for the form factor in $p^2$ and $q_2^2$. 

The form factors $H_S$ and $H_P$ are given by the diagrams of Fig \ref{fig:Hst}. 

\begin{figure}[h!]
\centering
\begin{tabular}{cccc}
\includegraphics[width=3.5cm]{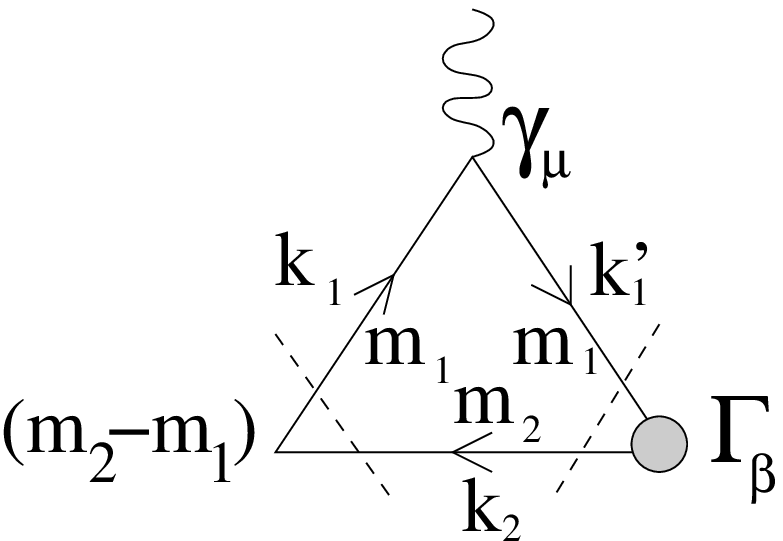} &&&
\includegraphics[width=3.5cm]{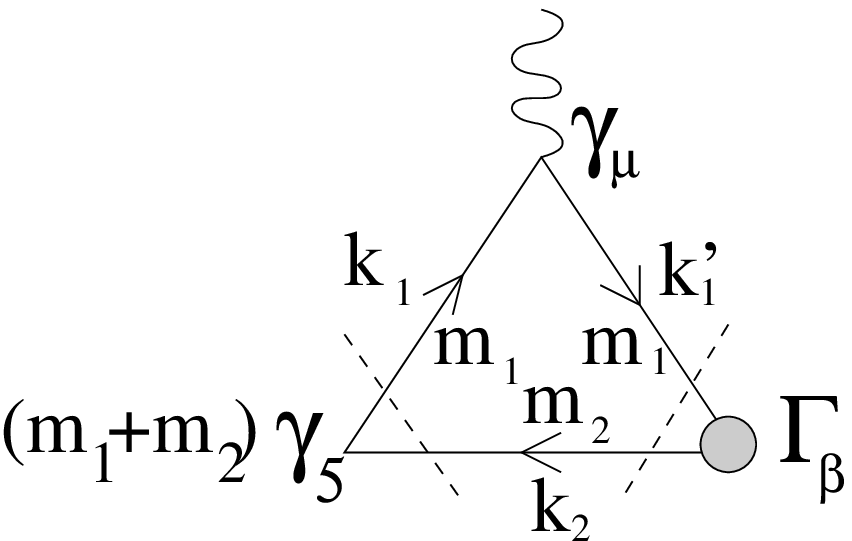} 
\end{tabular}
\caption{\label{fig:Hst} The triangle diagrams for $H_S^{(1)}(m_1,m_2)$ and $H_P^{(1)}(m_1,m_2)$.}
\end{figure}

Each diagram contains the contributions of photon emission both from the light and the heavy quarks:
  
\begin{eqnarray}
\begin{array}{ll}
H_S=Q_d H_S^{(d)}-Q_c H_S^{(c)}, &
H_P=Q_dH_P^{(d)}+Q_cH_P^{(c)}. 
\end{array}
\end{eqnarray} 

The corresponding single dispersion integrals has the form 
\small
\begin{eqnarray}
\label{hsdisp}
H_S^{(1)}(m_1,m_2)&=&\frac{\sqrt{N_c}}{4\pi^2}\int\limits_{(m_1+m_2)^2}^\infty
\frac{ds\;\phi_V(s)}{(s-p^2-i0)}(m_2-m_1)
\left(\rho_+(s,m_1,-m_2)+\frac{2\sqrt{s}}{p^2-M_V^2}\rho_{k_\perp^2}(s,m_1,m_2)\right), \\
H_P^{(1)}(m_1,m_2)&=&\frac{\sqrt{N_c}}{4\pi^2}\int\limits_{(m_1+m_2)^2}^\infty
\frac{ds\phi_V(s)}{(s-p^2-i0)}
(m_1+m_2)\left(\rho_+(s,m_1,m_2)+\frac{\rho_{k_\perp^2}}{\sqrt{s}+m_1+m_2}\right),  
\end{eqnarray}
\normalsize

where $\rho_+(s,m_1,m_2)$ and $\rho_{k_\perp^2}(s,m_1,m_2)$ are determined earlier in (\ref{rhoplus}) and (\ref{rhokperp2}).

For the $B$-decays of interest, we need the value of the form factors $H_{P,S}(p^2,q_1^2=0)$ at $p^2=M_B^2$,
which lies above the threshold $(m_c+m_q)^2$. The spectral representations for $H_{P,S}(p^2=M_B^2)$ develop the imaginary parts 
which occur due to the quark-antiquark intermediate states in the $p^2$-channel. It should be emphasized that no 
anomalous cuts emerge in the double spectral representation at $q_1^2\le 0$ \cite{lms_triangle}. 
In all cases considered in this paper,  
the value of $p^2=M_B^2$ lies far above the region of resonances which occur in the quark-antiquark channel.  
Far above the resonance region local quark-hadron duality works well and the calculation of the imaginary part based on the 
quark diagrams is trustable. The imaginary part turns out to be orders of magnitude smaller than the real 
part of the form factor and for the practical purpose of the decay rate calculation may be safely neglected.

%==========================================================================================================

\section{\label{Sect5}Numerical results}
The derived spectral representations for the form factors allow one to obtain numerical predictions for the 
form factors of interest as soon 
as the parameters of the model -- the meson wave finctions and the quark masses -- are fixed.

The wave function $\phi_i(s)$, $i=P,V$ can be written as  
\begin{eqnarray}
\label{phi}
\phi_i(s)=
\frac{\pi}{\sqrt2}
\frac{\sqrt{s^2-(m_1^2-m^2)^2}}{\sqrt{s-(m_1-m)^2}}
\frac{w_i(k^2)}{s^{3/4}}, \qquad 
k^2=\lambda(s,m_1^2,m^2)/4s, 
\end{eqnarray}
with $w_i(k^2)$ normalized as follows 
\begin{eqnarray}
\label{norm}
\int w^2_i(k^2)k^2 dk=1. 
\end{eqnarray}
The meson weak transition form factors from dispersion approach reproduce correctly the structure of the 
heavy-quark expansion in QCD for heavy-to-heavy and heavy-to-light meson transitions, as well as for the 
meson-photon transitions, if the radial wave functions $w(k^2)$ are localized in a region of
the order of the confinement scale, $k^2\le\Lambda^2$ \cite{melikhov}. 
Following \cite{ms} we assume a simple gaussian parameterization of the radial wave function
\begin{equation}
\label{gauss}
w_i(k^2)\propto\exp(-k^2/2\beta_i^2),
\end{equation}
which satisfies the localization requirement for $\beta\simeq \Lambda_{QCD}$ and 
proved to provide a reliable picture of a large family of the transition form factors \cite{ms}. 

We use the same values of the constituent quark masses and the quark couplings as have been obtained in \cite{ms}
\begin{equation}
\label{quark_masses}
\quad m_d=m_u=0.23 \;{\rm GeV}, \quad m_s=0.35 \;{\rm GeV},\quad m_c=1.45 \;{\rm GeV},\quad m_b=4.85 \;{\rm GeV}.
\end{equation} 

We consider several annihilation-type $B$-decays which have the highest probabilities; the weak-annihilation 
quark diagrams which induce these decays are shown in Fig.~\ref{fig:quark_diag}. 

\begin{figure}[h!]
\centering
\begin{tabular}{c}
\includegraphics[width=12cm]{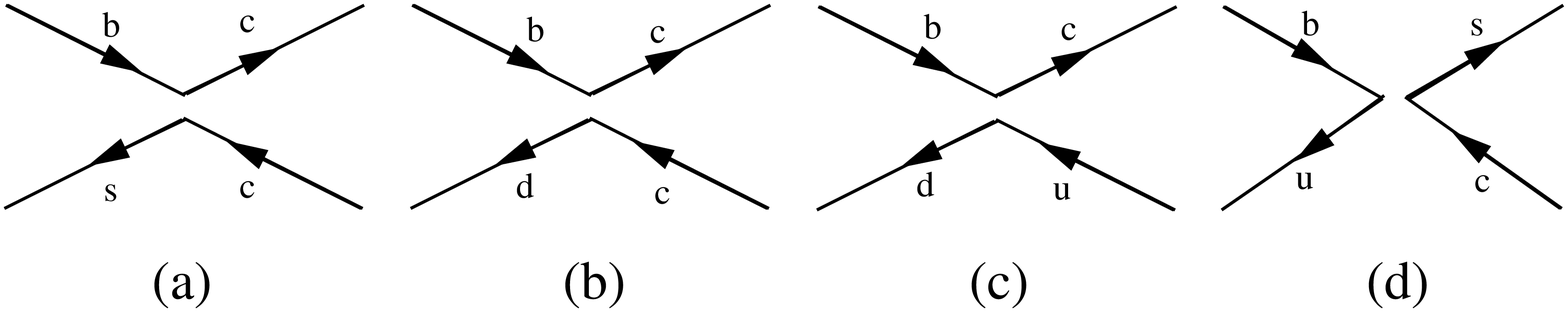} 
\end{tabular}
\caption{\label{fig:quark_diag}
Four-quark operators inducing the annihilation $B$-decays listed in Table \ref{table:wa-amplitudes}. 
(a) $\bar B^0_s\to J/\psi\gamma$; 
(b) $\bar B^0_d\to J/\psi\gamma$;  
(c) $\bar B^0_d\to D^{*0}\gamma$;  
(d) $B^{-}\to D_s^{*-}\gamma$. }
\end{figure}

The corresponding $F_{PC}$ and $F_{PV}$ and the decay rates are summarized in Table \ref{table:wa-amplitudes}. To highlight the 
contribution to the amplitudes coming from the photon emission from the $V$-meson loop, we multiply it by a coefficient $r$ 
which is set to unity in the decay-rate calculations. 
Obviously, for some modes the photon emission from the vector-meson loop is comparable or even exceeds  
the photon emission from the $B$-meson loop and thus should be taken into account. 

\begin{table}[h!]
\caption{\label{table:wa-amplitudes} 
The amplitudes and the branching ratio for the annihilation-type decay of $B$ and $B_s$.}
\centering
\begin{tabular}{|c|c|c|c|c|}
\hline
Reaction               & CKM-factor            &    $F_{PC}$ [GeV]   &    $F_{PV}$ [GeV]&       Br       \\
\hline
$\bar B^0_s\to J/\psi\gamma$  &  $a_2 V_{cb}V_{cs}^*$  &   $0.036-0.052 r$   &       0.020     &   $1.43\cdot 10^{-7}\left(\frac{a_2}{0.15}\right)^2$  \\
$\bar B^0_d\to J/\psi\gamma$  &  $a_2 V_{cb}V_{cd}^*$  &   $0.035-0.050 r$   &       0.021     &   $7.54\cdot 10^{-9}\left(\frac{a_2}{0.15}\right)^2$  \\
$\bar B^0_d\to D^{*0}\gamma$  &  $a_2 V_{cb}V_{ud}^*$  &   $0.012-0.014 r$ &   $0.007+0.002 r$ &   $4.33\cdot 10^{-8}\left(\frac{a_2}{0.15}\right)^2$  \\
$B^{-}\to D_s^{*-}\gamma$  &  $a_1 V_{ub}V_{cs}^*$  &  $-0.025+0.001 r$ & $-0.014+0.002 r$     &  $1.68\cdot 10^{-7}\left(\frac{a_1}{1.02}\right)^2$  \\
\hline
\end{tabular}
\end{table} 
For the scale-dependent Wilson coefficients $C_i(\mu)$ and $a_{1,2}(\mu)$ at the renormalization scale $\mu\simeq 5$ GeV 
we use the following values \cite{heff}: $C_1=1.1$, $C_2=-0.241$, $a_1=C_1+C_2/N_c=1.02$, and $a_2=C_2+C_1/N_c=0.15$. 
Similar values are used for numerical etimates in \cite{ch2004}: e.g., for $B_{(s)}\to J/\psi\gamma$ decay, 
$a_2=0.15$ in our analysis corresponds to the effective Wilson coefficient $\bar a_q=0.163$.

%===========================================================================================================

\section{\label{Sect6}Conclusion}
We have analyzed the annihilation-type radiative $B$ decays in the factorization approximation, 
taking into account both the photon emission both from the $B$-meson loop and the vector-meson loop ($V$-loop).  We have shown that 
the photon emission from the $V$-loop is not suppressed compared to the photon emission from the $B$-loop and gives a 
comparable contribution. 
We obtain predictions for the annihilation-type decays with the largest branching fractions: 
\begin{eqnarray}
{\cal B}(\bar B^0_s\to J/\psi\gamma)    &=&1.43\cdot 10^{-7}\left(\frac{a_2}{0.15}\right)^2, \\
{\cal B}(\bar B^0_d\to J/\psi\gamma)    &=&7.54\cdot 10^{-9}\left(\frac{a_2}{0.15}\right)^2, \\
{\cal B}(\bar B^0_d\to D^{0*}\gamma)    &=&4.33\cdot 10^{-8}\left(\frac{a_2}{0.15}\right)^2,  \\
{\cal B}(B^-\to \bar D_s^{*-}\gamma)    &=&1.68\cdot 10^{-7}\left(\frac{a_1}{1.02}\right)^2.
\end{eqnarray}
We would like to emphasize a relatively large branching ratio of the $\bar B^-\to \bar D_s^{*-}\gamma$ decay which 
makes this mode a prospective candidate for the experimental studies in the near future. 

In conclusion, we believe that in comparison with the existing estimates \cite{ch2004,ch2006} credit should be given to our results: 
First, we take into account those contributions which have been neglected previously but which are shown to give sizeable 
contributions to the amplitude. Second, our calculation of the form factors is based on a 
more detailed model for the $B$-meson structure. 

%===========================================================================================================
\newpage
\vspace{.5cm}
\appendix{\bf Acknowledgments} 

The author thank Dmitri Melikhov and Nikolai Nikitin for stimulating discussions on the subject of this talk. The work was supported by a Grant for Leading Scientific Schools 3042.2014.2 (Russia).  

%===========================================================================================================
\bibliographystyle{aipproc}

\begin{thebibliography}{99}
\bibitem{Kozachuk:2015kos} 
  A.~Kozachuk, D.~Melikhov and N.~Nikitin,
  %``Annihilation type rare radiative $B_{(s)}\to V\gamma$ decays,''
  arXiv:1511.03540 [hep-ph].
\bibitem{BaBar2004} 
B.~Aubert {\it et al.} [BaBar Collaboration],
%``Search for the decay $B^0 \to J/\psi \gamma$,''
Phys.\ Rev.\ D {\bf 70}, 091104 (2004).
% [hep-ex/0408018].
\bibitem{LHCb2015}
R.~Aaij {\it et al.} [LHCb Collaboration]  	
%``Search for the rare decays $B^0\to J/\psi\gamma$ and $B^0_s\to J/\psi\gamma$'', 
arXiv:1510.04866[hep-ex]. 
\bibitem{grinstein}
B.~Grinstein and R.~Lebed, Phys.~Rev.~D {\bf 60}, 031302 (1999).
\bibitem{gp}  
B.~Grinstein and D.~Pirjol, Phys. Rev. D {\bf 62}, 093002 (2000). 
%\bibitem{eeg}
%J.~A.~Macdonald Sorensen and J.~O.~Eeg, 
%``On the decay modes $B\to D*\gamma$''
%Phys.\ Rev.\ D {\bf 75}, 034015 (2007). 
\bibitem{ch2004} 
Y.~D.~Yang, G.~Lu, and R.~Wang,
%``The Rare radiative annihilation decays anti-B0(s,d) ---> J / psi gamma,''
Eur.\ Phys.\ J.\ C {\bf 34}, 291 (2004).
% [hep-ph/0308256].
\bibitem{ch2006} 
Y.~Li and C.~D.~L\"u,
%``Annihilation Type Radiative Decays of $B$ Meson in Perturbative QCD Approach'', 
Phys.\ Rev.\ D {\bf 74}, 097502 (2006).
% [hep-ph/0605220].
\bibitem{m}
D.~Melikhov, 
%``Dispersion approach to quark binding effects in weak decays of heavy mesons''
Eur.~Phys.~J.~direct {\bf C4}, 2 (2002) [hep-ph/0110087]. 
\bibitem{melikhov} 
D. Melikhov, Phys. Rev. D {\bf 53}, 2460 (1996); D {\bf 56}, 7089 (1997).
\bibitem{ms} 
M. Beyer and D. Melikhov, Phys. Lett. B {\bf 451}, 121 (1999), \\
D. Melikhov and B. Stech, Phys. Rev. D {\bf 62}, 014006 (2000). 
\bibitem{bmns} 
M.~Beyer, D.~Melikhov, N.~Nikitin, and B.~Stech,  Phys. Rev. D {\bf 64}, 094006 (2001). 
\bibitem{mn2004} 
D.~Melikhov and N.~Nikitin,
%``Rare radiative leptonic decays B(d,s) ---> l+l- gamma,''
Phys.\ Rev.\ D {\bf 70}, 114028 (2004). %[hep-ph/0410146].
\bibitem{heff} 
G.~Buchalla, A.~J.~Buras, and M.~E.~Lautenbacher, Rev.~Mod.~Phys.~{\bf 68}, 1125 (1996). 
\bibitem{mk} 
F.~Kruger and D.~Melikhov,
%``Gauge invariance and form-factors for the decay B ---> gamma l+ l-,''
Phys.\ Rev.\ D {\bf 67}, 034002 (2003).
%  [hep-ph/0208256].
\bibitem{amn}
V. V. Anisovich, D. Melikhov, V. Nikonov, 
Phys. Rev. D {\bf 52}, 5295 (1995); Phys. Rev. D {\bf 55}, 2918 (1997).  
\bibitem{korch} 
G. Korchemsky, D. Pirjol, and T.-M. Yan, Phys. Rev. D {\bf 61}, 114510 (2000). 
\bibitem{leet} 
J. Charles {\it et al.}, Phys. Rev. D {\bf 60}, 014001 (1999). 
\bibitem{lms_triangle} 
W.~Lucha, D.~Melikhov, and S.~Simula, 
%``Dispersion representations and anomalous singularities of the triangle diagram'', 
Phys.~Rev.~D {\bf 75}, 016001 (2007), Phys.~Rev.~D {\bf 92}, 019901 (2015). 
%======================================================================================parameters
\bibitem{pdg}
K.~A.~Olive {\it et al.} (Particle Data Group), Chin.~Phys.~C {\bf 38}, 090001 (2014).
\bibitem{lmss}
W.~Lucha, D.~Melikhov, H.~Sazdjian, and S.~Simula,
%``Strong three-meson couplings of $J/\psi$ and $\eta_c$'',  
arXiv:1506:09213.   
\bibitem{lmsfB}
W.~Lucha, D.~Melikhov, and S.~Simula, 
%``Decay constants of heavy pseudoscalar mesons from QCD sum rules''
J.~Phys.~{\bf G38}, 105002 (2011); 
%''Accurate bottom-quark mass from Borel QCD sum rules for fB and fBs'' 
Phys.~Rev.~D {\bf 88}, 056011 (2013);  
%``Accurate decay-constants ratios fB*/ fB and fBs*/ fBs from Borel QCD sum rules''
Phys.~Rev.~D {\bf 91}, 116009 (2015).   
\bibitem{lmsfD*}
W.~Lucha, D.~Melikhov, and S.~Simula, Phys.~Lett.~B {\bf 735}, 12 (2014).
%\bibitem{fDDs*lat}
%D.~Becirevic {\it et al.}, JHEP {\bf 1202}, 042 (2012).
\bibitem{fpsi_lat}
G.~C.~Donald {\it et al.}, Phys.~Rev.~D {\bf 86}, 094501 (2012).
\bibitem{mn_oscillations}
D.~Melikhov and N.~Nikitin, 
%``Nonfactorizable effects in the B - anti-B mixing''
Phys.~Lett. B {\bf 494}, 229 (2000).
\bibitem{bbns}
M.~Beneke, G.~Buchalla, M.~Neubert, and C.~T.~Sachrajda, 
%``QCD factorization for $B\to\pi\pi$ decays: Strong phases and CP violation in the heavy quark limit''
Phys.~Rev.~Lett. {\bf 83}, 1914 (1999). 
\end{thebibliography}

\end{document}